# Resilient Architectures for Free Space Optical Wireless Interconnection Systems

Sanaa Hamid Mohamed[a], Osama Zwaid Alsulami[a], Taisir E. H. El-Gorashi[a], Mohammed T. Alresheedi[b], and Jaafar M. H. Elmirghani*[a]

[a] School of Electronic and Electrical Engineering, University of Leeds, LS2 9JT, United Kingdom
[b] Department of Electrical Engineering, King Saud University, Riyadh, Kingdom of Saudi Arabia

**ABSTRACT**

In this paper, we propose the use of two Passive Optical Network (PON)-based network architectures to connect free-space Optical Wireless Communication (OWC) Access Points (APs) within a room with multiple users. We optimize through a Mixed Linear Integer Programming (MILP) model the assignment of mobile OWC users to more than one AP to improve the resilience of the fronthaul network, i.e the OWC system and the wired network linked to APs, and study the impact of users distribution and channel characteristics.

## 1. INTRODUCTION

Optical wireless communication (OWC) systems can offer a solution to the spectrum crunch in access networks[1,2]. The Radio Frequency (RF) spectrum is becoming increasingly exhausted as the traffic is currently continuing to increase at 30% to 40% per year leading to expectations of traffic growth by 30x in 10 years and by 1000x in 20 years[3]. It was reported that about 80% of all Internet traffic starts and terminates indoors[4]. Thus, efficient and high capacity designs for wired and wireless access networks inside building are required to meet these indoor demands. Optical wireless free space systems offer numerous advantages including abundant unregulated optical spectrum, freedom from fading, added physical layer security as light does not penetrate opaque objects, and the ability to re-use the same wavelengths in adjacent rooms. Various receiver and transmitter design considerations were studied to improve the performance of Infrared OWC systems by reducing the impact of noise, interference, and multipath dispersion while accounting for eye safety[5-21]. These include the use of different modulation formats and coding schemes[5-8], fly eye receivers with multiple detectors[6-11], spot diffusing transmitters with angle diversity receivers[12-16], imaging receivers[17], with delay adaptation methods[18] and with relay nodes[19], and the use of adaptive holograms in transmitters with imaging receivers[20] or angle diversity receivers[21]. Those solutions can then provide up to 10 Gbps OWC systems[22-25]. Visible Light Communication (VLC) was also considered to achieve multi Gigabit OWC systems with multiple users[26-31]. For downlinks, VLC systems can use Red Yellow Green Blue (RYGB) Laser Diodes (LDs) in a number of distributed Access Points (APs) for lighting and communication. The assignment of wavelengths and access points to users can be optimized while accounting for the users' distribution and indoor channel characteristics[2,32].

For demands that originate and terminate indoor such as user-to-user communication and for user-to-Internet traffic, improving the backhaul network (i.e. the connections between the APs within the room and connections between the APs and gateways) can reduce data transmission delay and networking power consumption, which is beneficial for users and service providers, and can increase the resilience of the system against AP failures. Links between APs located within a room can be wired or wireless while linking APs within a building requires a wired network. Wang et. al. proposed a Wavelength Division Multiplexing (WDM)-based OWC system that integrated a Central Office (CO) equipment with a fibre distribution network inside the building[33]. Kazemi et.al. proposed and optimized a multi-hop optical wireless backhaul network to link a number of optical attocell networks in a multi-tier manner to form a super cell with a single-gateway[34]. In this work, we propose the use of two Passive Optical Network (PON)-based network designs, a WDM-based architecture and a point-to-point architecture, to connect the APs within a building and to provide connections for the APs with a central office. For the wireless part of this access network design, we consider a WDM OWC downlink system with multiple APs and Angle Diversity Receivers (ADR)[2]. To provide resilient services, we propose assigning users to more than one AP and we use a Mixed Integer Linear Programming (MILP) model to optimize users assignments. We use the model to evaluate the performance and resilience of the OWC system against AP failure and consider the impact of user distribution and optical wireless channel characteristics. We also provide a preliminary evaluation of the resilience of the two proposed backhaul networks compared to switch-based data network architectures.

*j.m.h.elmirghani@leeds.ac.uk ; phone +44(0)113 343 2013; leeds.ac.uk

The remainder of this paper is organized as follows: Section 2 describes the resilient access network system model, the parameters considered, technologies used in the OWC system and the backhaul network and provides the optimization model. Section 3 provides the results while Section 4 presents the conclusions and future work.

## 2. RESILIENT ACCESS NETWORK MODEL

This section describes the WDM OWC system with multiple APs and users, provides a brief overview of the resource allocation optimization MILP model and introduces the two proposed PON-based data networks for connecting the access points within a room.

### 2.1 The Optical Wireless Communication System

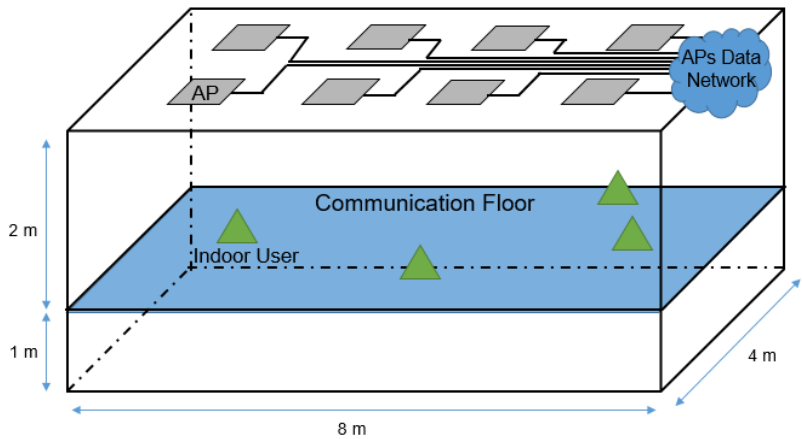

Figure 1.System model: A 3m height room with multiple users and 8 WDM access points (AP) in the ceiling. The users are distributed in a communication floor 1 m above the room floor. The APs are interconnected through a data network.

The room setup we used to model the OWC system[2] is shown in Figure 1. The room is 3m heigh with 8 access points in the ceiling and the communication floor is 1m above the room floor. Each access point provides WDM downlinks by utilizing LDs with four wavelength, i.e. RYGB LDs, which are in the visible light range, hence, can be used for communication and illumination. The parameters of the room, users, transmitters in the APs and the receivers are summarized in Table 1. An ADR with 4 branches is used as the receiver. These parameters were used to ray trace[2,27,35] and simulate all possible transmitted beams (i.e. $8 \times 4 \times 4$) with up to second order reflections from the APs to 32 predefined and equally distributed points in the communication floor[2]. Also, similar evaluation for the background noise was performed.

Up to 7 users are considered with uniform random allocation in one of the predefined 32 locations. Prior knowledge for the number of users and their locations is assumed to be provided by a controller connected to all APs[32]. The APs are connected through a data network that allows direct connections between the APs and also provides connections with central office equipment such as an optical line terminal (OLT) to provide connections to the Internet. Selecting a high bandwidth data network with multiple paths between the APs can enable direct user-to-user data transmission without the need for indirect transmission through the Internet. Hence, better performance in terms of delay and power consumption can be achieved. We propose the use of two PON-based network architectures for connecting the APs. A brief review of these networks is presented in the following subsection.

Table 1. Room and OWC system configuration[2].

| Room and ray tracing parameters | |
| --- | --- |
| Parameter | Value |
| Length (x), width (y), height (z) | 8m, 4m, 3m |
| Reflection coefficient for walls and ceiling | 0.8 |
| Reflection coefficient for floor | 0.3 |
| Number of reflections/area of reflecting element | 1/5cm×5cm , 2/20cm×20cm |
| Order of Lambertian pattern for all surfaces | 1 |
| Semi-angle of reflection element at half-power | 60° |
| **Users** | |
| Number of users | Up to 7 users |
| Height | 1 m |
| Distribution | Uniform random in 32 locations |
| **APs (transmitters)** | |
| Number | 8 |
| (x,y,z) location | (1m,1m,3m),(1m,3m,3m), (1m,5m,3m),(1m,7m,3m), (3m,1m,3m),(3m,3m,3m), (3m,5m,3m),(3m,7m,3m) |
| Optical power of red, yellow, green, and blue LDs | 0.8 W, 0.5 W, 0.3 W, and 0.3 W |
| Semi-angle at half-power | 60° |
| Number of RYGB LDs per AP | 12 |
| **Receivers** | |
| Number of photodetectors | 4 |
| Azimuth/elevation angles of photodetectors | 45°/70°, 135°/70°,225°/70°,315°/70° |
| Field of view (FOV) of a photodetector | 25° |
| Area of a photodetector | 20 mm$^2$ |
| Responsivity for red, yellow, green, and blue | 0.4 A W$^{-1}$, 0.435 A W$^{-1}$ 0.3 A W$^{-1}$, 0.2 A W$^{-1}$ |
| Receiver bandwidth | 1.75 GHz |
| Receiver noise current spectral density | 4.47 pA/√Hz |

In this work, we propose resilient downlink connections, where each user can connect to more than one AP. We assume the use of On-Off Keying (OOK) and Forward Error Correction (FEC) and allow connections with Signal-to-Noise-and-Interference (SINR) more than or equal to 13.8 dB which provides Bit Error Rate (BER) of $1\times10^{-4}$. With an FEC code such as RS(255 239), and with an input BER of $1\times10^{-4}$, an output BER of $1\times10^{-14}$ is produced with a small overhead of 6%. We utilize a MILP model[2] to optimize the assignment of users to one or more WDM APs. For detailed derivations and linearization of non-linear terms in the MILP used to calculate the SINR, the reader is referred to our previous work[2]. It is worth noting that the objective function in our previous work was to maximize the sum of user SINR values and that each user is assigned to only one AP. In this work, we optimize the assignment of users to one or more APs and consider an objective function to increase the users assignments. The following provides a brief summary of the model used including key sets, parameters and variables, the objective function, and key constraints.

Sets:
$U$      Set of users.
$A$      Set of APs.
$W$     Set of wavelengths.
$B$     Set of receiver branches.

Parameters:
$R_{uf}^{a\lambda}$   The squared electrical current due to received optical power of user $u$ at receiver branch $f$ from AP $a$ using wavelength $\lambda$; $u \in U, f \in B, a \in A, \lambda \in W$.
$N_{uf}^{a\lambda}$   The background light noise mean square current received for user $u$ at receiver branch $f$ from AP $a$ using wavelength $\lambda$; $u \in U, f \in B, a \in A, \lambda \in W$.
$\sigma$      Mean square receiver noise current = Receiver noise power spectral density in A²/Hz times the receiver bandwidth.
$K$     Weighting factor =1000.
$Z$     SINR value that guarantee a BER of $1\times10^{-4}$.

Variables:
$\gamma_{uf}^{a\lambda}$   The SINR of user $u$ at receiver branch $f$ when receiving from AP $a$ using wavelength $\lambda$; $u \in U, f \in B, a \in A, \lambda \in W$.
$S_{uf}^{a\lambda}$   Binary variable that is equal to one if user $u$ at receiver branch $f$ is receiving from AP $a$ using wavelength $\lambda$, and is equal to zero otherwise; $u \in U, f \in B, a \in A, \lambda \in W$.

The SINR of user $u$ at receiver branch $f$ when receiving from AP $a$ using wavelength $\lambda$ can be expressed as[2]:

$$\gamma_{uf}^{a\lambda} = \frac{R_{uf}^{a\lambda} S_{uf}^{a\lambda}}{\sum_{b \in A, b \neq a} \sum_{m \in U, m \neq u} \sum_{g \in B} R_{uf}^{b\lambda} S_{mg}^{b\lambda} + \sum_{b \in A, b \neq a} N_{uf}^{b\lambda} [1 - \sum_{m \in U, m \neq u} \sum_{g \in B} S_{mg}^{b\lambda}] + \sigma}, \quad (1)$$

The term in the numerator represents the received signal while the first term in the denominator represent the interference, which is the sum of all received signals from other APs at the same wavelength when used for communication. The second term in the denominator represents the background light noise, which is the sum of all received signals from other APs at the same wavelength when they are only used for illumination (i.e. not assigned to any user)[2]. The objective is to maximize the SINR and the total number of assignments for all users to increase the system resilience against APs failures:

$$Maximise: \sum_{u \in U} \sum_{f \in B} \sum_{a \in A} \sum_{\lambda \in W} \left[ \gamma_{uf}^{a\lambda} + K S_{uf}^{a\lambda} \right]. \quad (2)$$

Subject to the following allocation constraints:

$$\sum_{\lambda \in W, f \in B} S_{uf}^{a\lambda} \leq 1, \forall\ u \in U, a \in A, \quad (3)$$

$$\sum_{u \in U, f \in B} S_{uf}^{a\lambda} \leq 1, \forall\ \lambda \in W, a \in A, \quad (4)$$

$$\gamma_{uf}^{a\lambda} \geq Z S_{uf}^{a\lambda}, u \in U, f \in B, a \in A, \lambda \in W. \quad (5)$$

Equation (3) ensures that each user can be assigned only one wavelength in an AP. Equation (4) ensures that each wavelength within an AP is only allocated once. Equation (5) ensures that a user is allocated only if the SINR will guarantee a BER of $1\times10^{-4}$.

This model takes the number of users and their locations information and the channel characteristics, and provides the optimum allocation results of users to branches in the receiver, APs, and wavelengths and the corresponding SINR values for the users where a user can be allocated to more than one AP.

## 2.2 Proposed Access Points Data Networks

Traditional Time Division Multiplexing (TDM) Passive Optical Networks (PONs) and traditional WDM PONs have been used to provide reliable, high performance and low maintenance and low power consumption connections with end users at the last mile of information and communication networks[37]. Using these passive networks in addition to three improved architectures was proposed for connecting end points in data centre networks with arbitrary and high bandwidth requirements[38]. The improved designs increase the network bandwidth and were proposed as the traditional TDM and WDM PONs were more suitable for lower end user to end user demands and asymmetric upload and download bandwidth requirements.

In this work and to improve the resilience of connecting the APs of the OWC system with the remaining access network (i.e. the OLT and higher network levels), we propose using two of the improved PON-based data networks to connect the APs. These designs are namely an Arrayed Waveguide Grating Router (AWGR)-based design that uses TDM/WDM[39-41], and a TDM or WDM point-to-point-based design that uses end points (i.e. the APs in this work) to forward traffic[42]. Both designs can provide multipath connections between any end point pairs. Improving the connectivity between the APs in terms of the capacity and multipath availability reduces the delay for any AP to AP or user to user communication and also improves the resilience of the whole access network for users within a room or a building.

Figure (2) shows an example of an AP data network that uses the AWGR-based PON design. In this example, we connect the 8 APs in a room together and with the OLT in the CO by using only passive components. Two AWGRs are proposed to provide non-blocking all-to-all connections[39] between all sets of APs and with the OLT. In the example provided in Figure (2), the APs are clustered into four sets. A proposed connections can use two 4×4 AWGRs and will require four wavelengths to provide the connections of each set with the OLT and the connections between different AP sets[38]. Each 2 APs in a set connect to a coupler/splitter that connects to a unique input/output port of one of the two AWGRs. The OLT is connected to both AWGRs to provide higher capacity for the user to OLT traffic (i.e. user to Internet traffic and vice versa). The remaining ports in the AWGRS provide multi hop connections between elements connected to different AWGRs. The additional splitters and Arrayed Waveguide Grating (AWG) ensure proper multiplexing of wavelengths. Each set of APs also has a passive and direct all-to-all network[38] to connect all the APs.

The routing and wavelength assignment to realize this design was detailed in previous work[38,39]. Each AP is equipped with a tuneable transceiver and can tune the wavelength to communicate with other APs in other sets or with the OLT. With 10 Gbps tuneable transceivers, a bisection bandwidth of 5×4×10=200 Gbps is achievable.

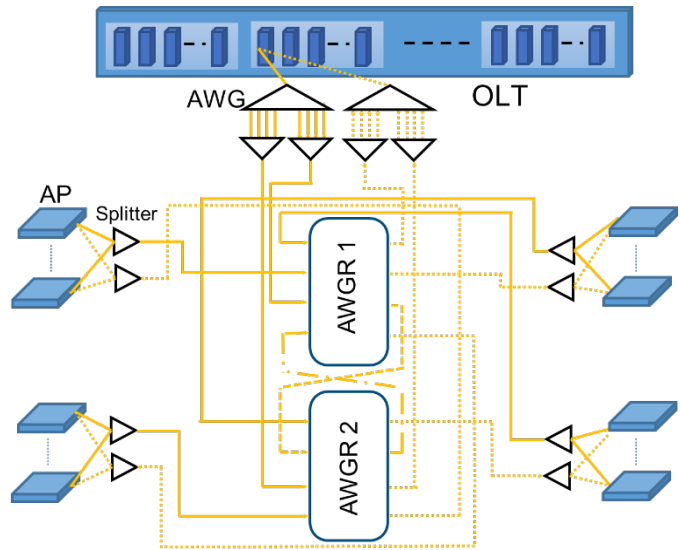

Figure 2. Connecting the access points in a room with the AWGR-based PON data network.

Figure (3) shows an example of the point-to-point PON design[42] that can be used to connect the APs within a room or within a building. In this design, APs are first clustered into groups, and each group is clustered into subgroups. APs within each subgroup connect to the remaining network through a splitter/coupler. Each subgroup connects to another subgroup in each group. Each group has a dedicated subgroup to connect with the OLT. The APs within a group are to be passively interconnected to provide all-to-all communication[38]. Each AP has a forwarding mechanism (e.g. by a Network Interface Card (NIC)) to enable transmitting other APs demands. Single wavelength, hence, only TDM, or multiple wavelengths can be used in the design. Due to the large number of paths between any AP pairs, higher resilience can be achieved against APs failures.

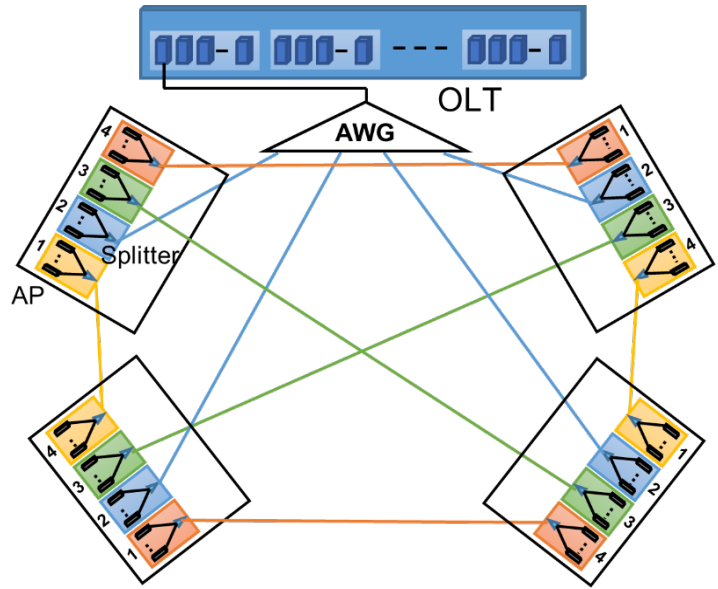

Figure 3. Connecting the access points within a room/building with the point-to-point PON data network.

## 3. RESULTS

This section provides the MILP-based optimized resource allocation results for the downlink OWC system when each user is able to connect to multiple APs. It also provides the optimum allocation results for the cases when one or two APs are not available.

### 3.1 Optimized resources allocation with multiple AP links

We first show the average SINR results of all the users when each user is allowed to connect with one AP only. The choice of AP is optimized[2]. Then, we show the allocation optimization results when multiple APs can be assigned to each user. These results are generated for 20 different and uniform random user locations (called allocations of the users) in each considered case for up to 7 users. Figure (4) and Figure (5) show the optimum allocation results when one and two users are present in the system, respectively. High SINR values between 18 dB and 19 dB are achievable in both cases when one AP is assigned to each user. This is due to the cases where the users locations allow assigning the user or the two users in interference-free allocations.

Figure (4) shows that when assigning a single user to more than one AP per user, the average SINR value for all users is reduced by about 2 dB. This is because allocating more links in the system introduces additional interference between the links that use the same wavelength. For the case of two users in the room as in Figure (5), the average SINR can drop by about 5 dB to a value around 13.8 dB, which is the lowest allowed SINR value, as the impact of the interference and background noise is higher due to the additional users compared to the results in Figure (4). In both cases, there is at least one assigned link for each user that provides a very high SINR value. Due to the limited FOV of each AP, the results in Figure 4(a) and Figure 5(b) show that a user cannot be allocated to more than 3 APs at the same time.

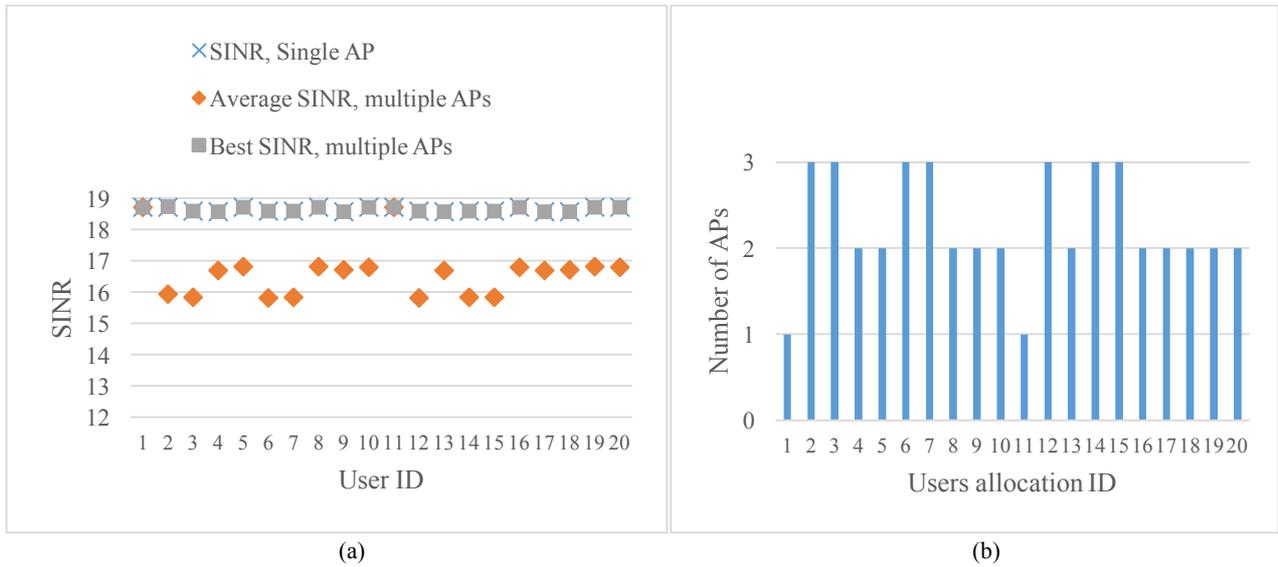

(a)                  (b)

Figure 4. Optimized resource allocation results for a single user: (a) SINR results, (b) Number of APs assigned to the user in each random allocation.

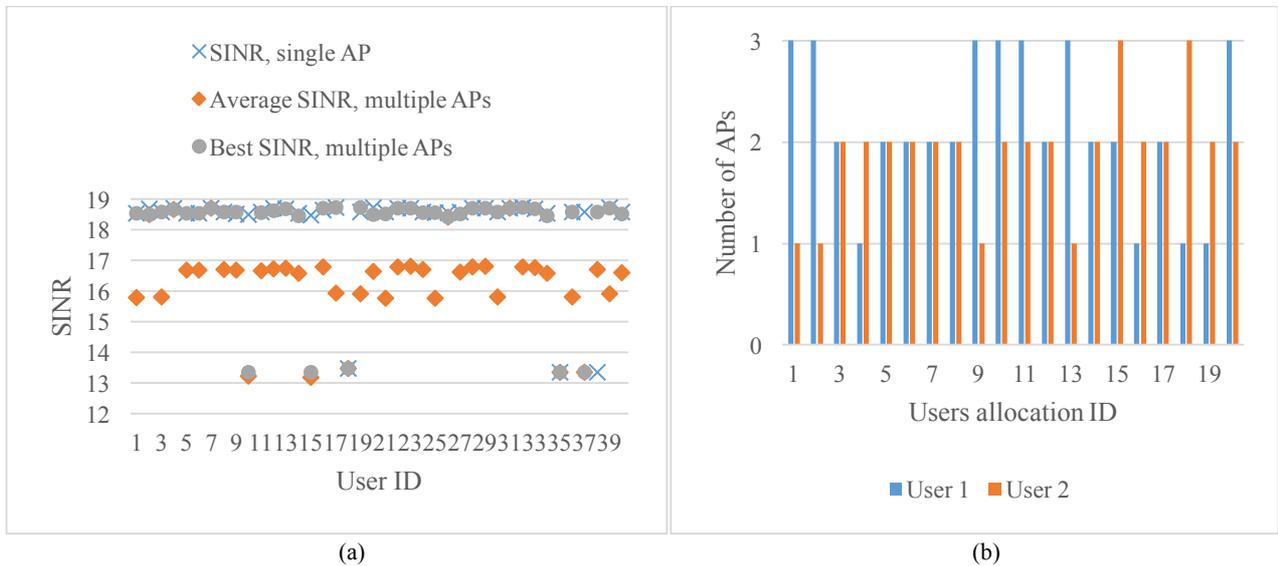

(a)                  (b)

Figure 5. Optimized resources allocation results for a two users: (a) SINR results, (b) Number of APs assigned to each user in each random allocation.

The average SINR results when allocating resources for up to 7 users in the room are provided in Figure (6). The results show that even for a high number of users, the OWC system is able to provide multiple APs connections to users while ensuring a BER value of $1\times10^{-4}$ for assigned users. Some of the allocation cases, for example for 3 users, examined some worst case scenarios, where according to the used objective function and constraints, it is more optimum to allocate some users to many APs while some users are not assigned to any AP. This has resulted an average SINR value below 13 dB. For the examined allocations with up to a maximum of 5 users in the room, the mode for the number of assigned APs is 2. For 6 and 7 users, the mode is 1.

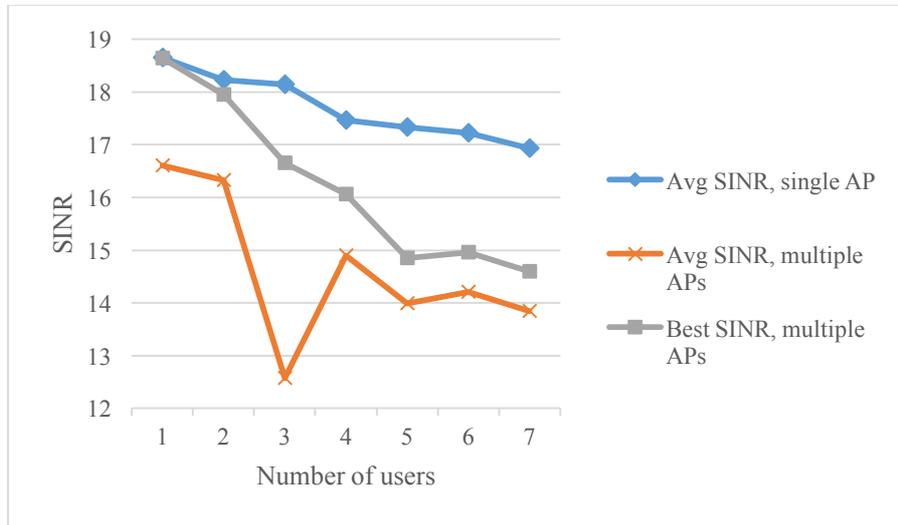

Figure 6. SINR results for the optimized resource allocation for up to 7 users.

### 3.2 Optimized resources allocation with AP failures

We evaluated the average SINR values based on the optimized resources allocation with three cases of AP failures. An AP can be inaccessible to a user either under hardware or software failures or if its optical beams are blocked by an object. The impact of optical beams blockage due to a disc, with several dimension parameters, between a user and APs was studies in our previous work[43]. The results show that when assigning two APs to a user, the maximum separation between the assigned APs results in reduced blockage.

In this work, we consider two cases of single AP failure and a case of two failures, which are randomly selected to be in access point 1, access point 5, and in both simultaneously. We use the same 20 uniform random allocations (ie user locations) for the users and evaluate the optimum resource allocation and average SINR results for 1, 2 and 3 users allocations. Figure (7) shows the average SINR results for the three cases and compares it to the average SINR results when all APs are available. The results show that any AP failure reduces the average SINR as either some of the users are not assigned to any APs due to unavailability of links with at least the minimum allowed SINR or due to the increased interference and background noise when optimizing the resources allocation with less available resources. The impact on average SINR reduction of a single AP failure depends on both the users locations and the inaccessible AP location (i.e. the channel characteristics), thus different AP failures can cause different average SINR results.

For the cases of single user and three users, the results show that the average SINR is worst with 2 access points failures compared with a single AP failure. This can be attributed to the increase in the interference and background noise when the available resources are lower, thus the results are less optimum in terms of the sum of SINR values, but more resilient assignments are selected. The results in the two users scenarios, shows a case of single AP failure (i.e. AP5) performing worst than 2 AP failures (i.e. AP1 and AP5). In these cases, more users are not assigned to any link, and hence, less interference and background noise for lower number of assigned link causes the average SINR value to be higher than the AP5 failure case.

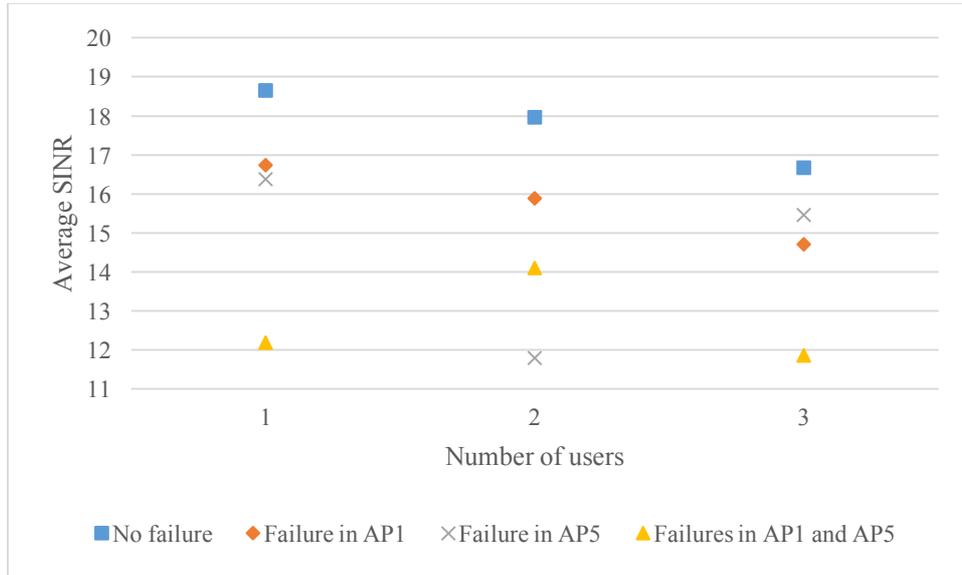

Figure 7. Average SINR results for the optimized resource allocation with AP failures.

Using the proposed AWGR-based PON or the point-to-point data networks, provides passive connections between each AP and the OLT, which improves the resilience of the access network. Using a switch-based data network[44] to connect the APs and improve user-to-user communication can worsen the resilience of the system. In these networks, a number of APs will be connected through an electronic switch to provide connections with other switches or the OLT. A failure in a switch is then equivalent to a failure in multiple APs.

## 4. CONCLUSIONS

In this paper, we proposed a resilient free space optical wireless system for multiple users that integrates the use of a visible light OWC system and PON based data networks. The OWC system utilizes RYGB LDs in multiple APs in the ceiling to provide downlink connections for indoor users with ADRs. This system allows each user to be assigned to more than one AP to improve the resilience against AP failures. We optimized the system through a MILP model considering the assignment of users to APs resources in cases of no AP failure and in three cases of AP failures and evaluated the impact of the optimized assignments on the average SINR values. Random allocations for the indoor users are considered to account for the indoor channel characteristics.

To improve the resilience of the access network and its performance for user-to-user traffic, we proposed the use of an AWGR-based PON or a point-to-point PON to passively connect the APs together and with the OLT. We showed that using a switch-based data network to connect the APs can worsen the resilience of the system as switch failures cause multiple APs disconnection whereas the proposed PONs improve the APs connections without reducing the resilience of the access network.

Future work includes considering angle diversity transmitters and imaging receivers in the WDM OWC system to improve the channel characteristics, evaluating and optimizing the average sum rate to improve the fairness between users in allocating AP resources and in resilience, considering Infrared downlinks, and evaluating the performance of the proposed PON data networks with realistic user-to-user indoor demands.

## ACKNOWLEDGEMENTS

This work was supported in part by the Engineering and Physical Sciences Research Council (EPSRC), in part by INTelligent Energy aware NETworks (INTERNET) under Grant EP/H040536/1, in part by SwiTching And tRansmission (STAR) under Grant EP/K016873/1, and in part by Terabit Bidirectional Multi-user Optical Wireless System (TOWS) project under Grant EP/S016570/1. All data is provided in the results section of this paper.